\documentclass[a4paper,conference]{IEEEtran}

\usepackage{cite}      % Written by Donald Arseneau

\usepackage{graphicx}  % Written by David Carlisle and Sebastian Rahtz

\usepackage{subfigure} % Written by Steven Douglas Cochran

\usepackage{fancyheadings}
\usepackage{listings}
\usepackage{amsfonts}
\usepackage{amsmath}
\usepackage{epsfig}
\usepackage{xspace}
\usepackage{flushend}
\newcommand{\oh}[1]
    {\mbox{$ {\mathcal O}( #1 ) $}}

\newcommand{\fig}[1]
    {Figure~\ref{fig:#1}}

\newcommand{\swift}
    {{\sc swift}\xspace}
\newcommand{\bsf}[1]
    {\textbf{\textsf{#1}}}

\begin{document}

\title{{\sc swift}: Fast algorithms for multi-resolution SPH on
multi-core architectures}

\author{
\IEEEauthorblockN{Pedro Gonnet\IEEEauthorrefmark{1}, Matthieu Schaller\IEEEauthorrefmark{2}, Tom Theuns\IEEEauthorrefmark{2}\IEEEauthorrefmark{3}, Aidan B. G. Chalk\IEEEauthorrefmark{1}}
\IEEEauthorblockA{\IEEEauthorrefmark{1}School of Engineering and Computing Sciences,\\
Durham University,\\
DH1 3LE Durham, UK\\ 
\{pedro.gonnet,aidan.chalk\}@durham.ac.uk \\
~\\}
\IEEEauthorblockA{\IEEEauthorrefmark{2}Institute for Computational Cosmology,\\
Durham University, \\
DH1 3LE Durham, UK\\ 
\{matthieu.schaller,tom.theuns\}@durham.ac.uk \\
~\\}
\IEEEauthorblockA{\IEEEauthorrefmark{3}Department of Physics, \\
 University of Antwerp, Groenenborgerlaan 171, \\
B-2020 Antwerp, Belgium}
}

%\specialpapernotice{(Invited Paper)}

\maketitle

\begin{abstract}
This paper describes a novel approach to neighbour-finding in
Smoothed Particle Hydrodynamics (SPH) simulations with large dynamic
range in smoothing length.
This approach is based on hierarchical cell decompositions, sorted interactions,
and a task-based formulation.
It is shown to be faster than traditional tree-based
codes, and to scale better than domain decomposition-based approaches on
shared-memory parallel architectures such as multi-cores.
\end{abstract}

%%%%%%%%%%%%%%%%%%%%%%%%%%%%%%%%%%%%%%%%%%%%%%%%%%%%%%%%%%%%%%%%%%%%%%%%%%%%%%%%
%  Introduction
%%%%%%%%%%%%%%%%%%%%%%%%%%%%%%%%%%%%%%%%%%%%%%%%%%%%%%%%%%%%%%%%%%%%%%%%%%%%%%%%
\section{Introduction}

Since the past few years, due to the physical limitations
on the speed of individual processor cores, instead of
getting {\em faster}, computers are getting {\em more parallel}.
This increase in parallelism comes mainly in the form of
{\em multi-core} computers in which the number cores can be
expected to continue growing exponentially, e.g.~following
Moore's law, much in the same way processor speeds were up until
a few years ago.

The predominant paradigm for parallel
computing is currently distributed-memory parallelism using MPI
(Message Passing Interface) \cite{ref:Snir1998},
in which large simulations are generally
parallelized by means of data decompositions, i.e.~by assigning
each node or core a portion of the data on which to work.
The cores execute the same code
in parallel, each on its own part on the data, intermittently exchanging data.
The amount of {\em computation} local to the node is then proportional
to the amount of data it contains, e.g.~its volume, while
the amount of {\em communication} is proportional to the
amount of computation spanning two or more nodes, e.g.~its
surface.
If the number of nodes increases, or 
smaller problems are considered, the surface-to-volume ratio,
i.e. the ratio of communication to computation,
grows, and the time spent on communication will increasingly
dominate the entire simulation, reducing scaling and parallel
efficiency.

Assuming the individual cores do not get any faster,
this means that small simulations, for which the
maximum degree of parallelism has already been reached, will never
become any faster.
In order to speed up small simulations, or to continue
scaling for large simulations, new approaches on how
computations are parallelized need to be considered.

With the above in mind, we will, in the following,
describe a reformulation of the
underlying algorithms for Smoothed Particle Hydrodynamics (SPH)
simulations which uses asynchronous
task-based shared-memory parallelism to achieve better parallel
scaling and efficiency on multi-core architectures.

%%%%%%%%%%%%%%%%%%%%%%%%%%%%%%%%%%%%%%%%%%%%%%%%%%%%%%%%%%%%%%%%%%%%%%%%%%%%%%%%
%  Algorithms, i.e. decomposition, interactions, and parallelization
%%%%%%%%%%%%%%%%%%%%%%%%%%%%%%%%%%%%%%%%%%%%%%%%%%%%%%%%%%%%%%%%%%%%%%%%%%%%%%%%
\section{Algorithms}

The interactions in compressible gas dynamics using SPH
are computed in two distinct stages that are
evaluated separately:
\begin{enumerate}
    \item {\em Density} computation: For each particle $p_i$,
        loop over all particles $p_j$ within $h_i$ of $p_i$ and 
        compute the particle densities.
    \item {\em Force} computation: For each particle $p_i$,
        loop over all particles $p_j$
        within $\max\{h_i,h_j\}$ and compute the particle forces.
\end{enumerate}
The identification of these interacting particle pairs
incurs the main computational cost,
as will be shown in the following sections,
and therefore also presents the main challenge in implementing efficient
SPH simulations.

\subsection{Tree-based approach}

In its simplest formulation, all particles in an SPH simulation have
a constant smoothing length $h$.
In such a setup, finding the particles in range of any other particle
is similar to Molecular Dynamics simulations, in which all particles
interact within a constant cutoff radius, and approaches which are used
in the latter, e.g. cell-linked lists
\cite{ref:Allen1989} or Verlet lists \cite{ref:Verlet1967} can be used
\cite{ref:Dominguez2011,ref:Viccione2008}.

The neighbour-finding problem becomes more interesting, or difficult,
in SPH simulations with variable smoothing lengths, i.e.~in which
each particle has its own smoothing length $h_i$, with ranges spawning
up to several orders of magnitude.
In such cases, e.g. in Astrophysics simulations \cite{ref:Gingold1977},
the above-mentioned approaches cease to work efficiently.
Such codes therefore usually rely on {\em trees}
for neighbour finding \cite{ref:Hernquist1989,ref:Springel2005,ref:Wadsley2004},
i.e.~$k$-d trees \cite{ref:Bentley1975} or octrees \cite{ref:Meagher1982}
are used to decompose the simulation space. 
The particle interactions are then computed by traversing the list of
particles and searching for their neighbours in the tree.

Using such trees, it is in principle trivial to parallelize
the neighbour finding and the actual computation on shared-memory
computers,
e.g.~each thread walks the tree for a different particle,
identifies its neighbours and computes the densities and/or
the second derivatives of the physical quantities for that particle.

Despite their simple and elegant formulation, the tree-based
approach to neighbour-finding has three main problems:
\begin{itemize}
    \item Computational efficiency: Finding all neighbours
        of a particle in the tree is, on average, in \oh{\log N},
        and worst-case behavior in \oh{N^{2/3}} \cite{ref:Lee1977},
        i.e.~the cost per particle grows with the
        total number of particles $N$.
    \item Cache efficiency: When searching for the neighbours of a
        given particle, the data of all potential neighbours, which may
        not be contiguous in memory, is traversed separately.
    \item Symmetry: The parallel tree search can not exploit symmetry,
        i.e.~a pair $p_i$ and $p_j$ will always be found twice,
        once when walking the tree for each particle.
\end{itemize}
    
These problems are all inherently linked to the use of
spatial trees, i.e. more specifically their traversal,
for neighbour-finding.
As of here we will proceed differently, using a hierarchical cell
decomposition, and computing the interactions by cell pairs,
avoiding the above-mentioned problems.

\subsection{Spatial decomposition}

If $h_\mathsf{max} := \max_i  h_i $ is the maximum smoothing
length of any particle in the simulation, we start by splitting
the simulation domain into rectangular cells of edge length
larger or equal to $h_\mathsf{max}$.

Given such an initial decomposition, we then generate
a list of cell {\em self-interactions}, which contains all
non-empty cells in the grid.
This list of interactions is then extended by the cell
{\em pair-interactions}, i.e. a list of all non-empty cell pairs
sharing either a face, edge, or corner.
For periodic domains, cell pair-interactions need also be
specified for cell neighbouring each other across
periodic boundaries.
These self- and pair-interactions encode, conceptually at least,
the evaluation of the interactions between all particles
in the same cell, or all interactions between particle pairs
spanning a pair of cells, respectively.

\begin{figure}
    \centerline{\epsfig{file=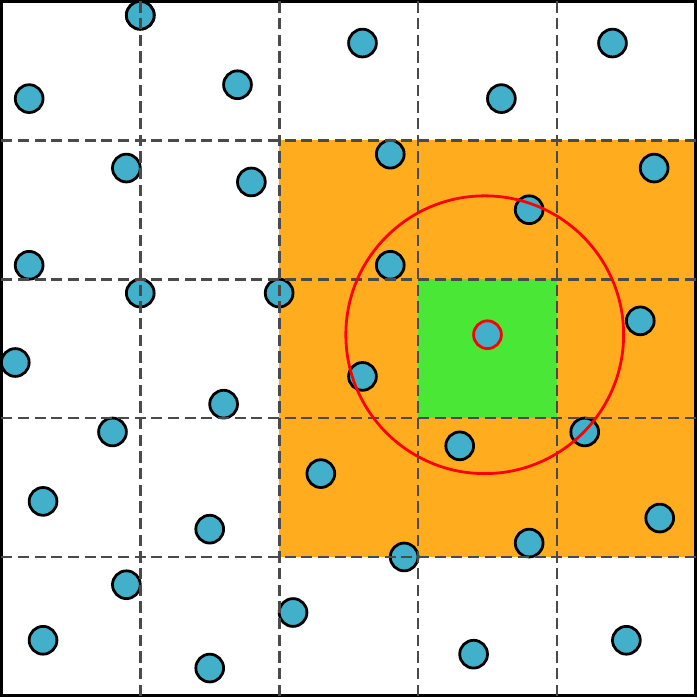,width=0.5\columnwidth}}
    
    \caption{Initial spatial decomposition: The simulation volume is divided into cells of
        edge length greater or equal to the largest smoothing length in the
        system. All neighbours of any given particle (small red circle) within
        that particle's smoothing length (large red circle) are guaranteed to lie
        either within that particle's own cell (green) or the directly
        adjacent cells (orange).}
    \label{fig:InitialDecomp}
\end{figure}

In this first coarse decomposition, if a particle $p_j$
is within range of a particle $p_i$, both will be either
in the same cell, or in neighbouring cells for which a
cell self-interaction or cell pair-interaction has been
specified respectively (see \fig{InitialDecomp}).

In the best of cases, i.e.~if each cell contains
only particles with smoothing length equal to the cell
edge length, if for any particle $p_i$ we inspect each
particle $p_j$ in the same
and neighbouring cells, only roughly 16\% of the $p_j$
will actually be within range of $p_i$ \cite{ref:Gonnet2007}.
and is even worse if the cells contain particles who's smoothing length
is less than the cell edge length.
We therefore recursively refine the cell decomposition,
bisecting each cell along
all spatial dimension if: (a) The cell contains more than
some minimal number of particles, and (b) the smoothing
length of a reasonable fraction of the particles within
the cell is less than half the cell edge length.

\begin{figure}
    \centerline{{\scriptsize\bsf A} \raisebox{-0.85\height}{\epsfig{file=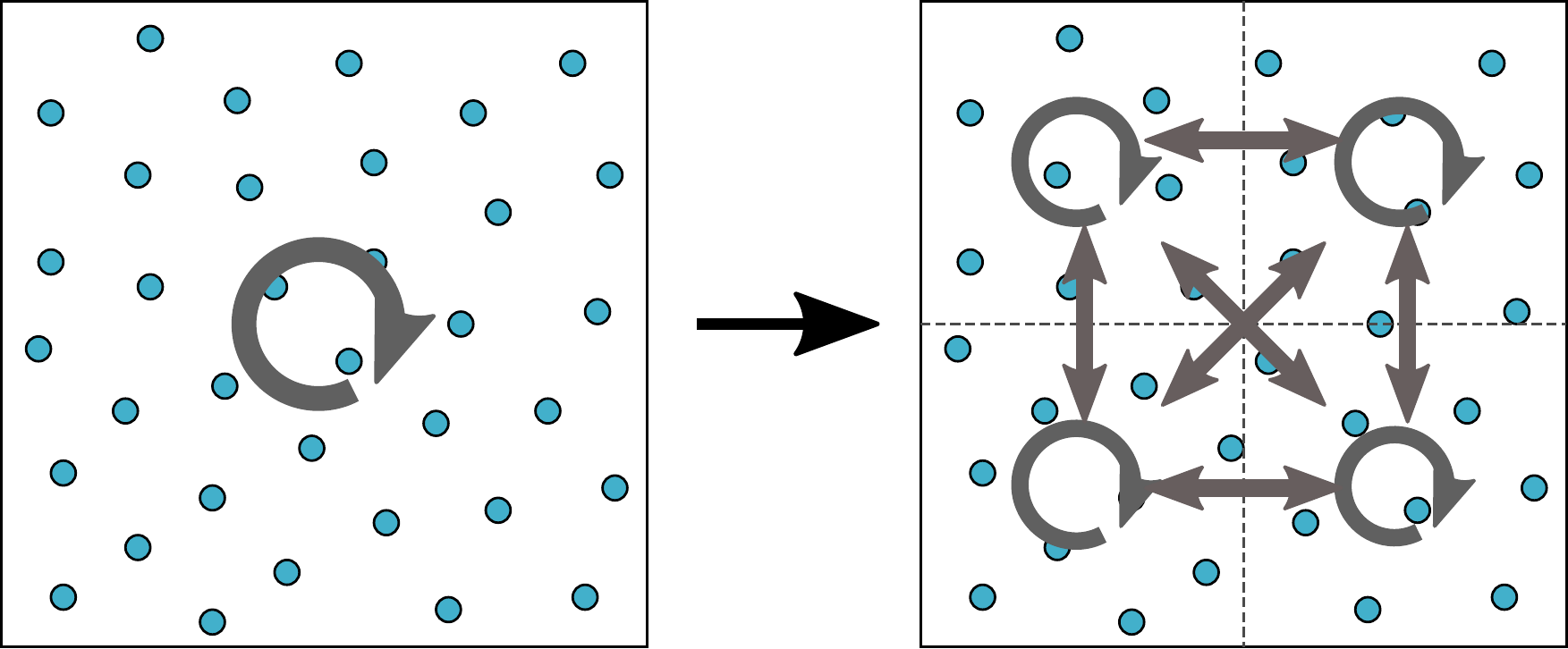,width=0.44\columnwidth}}}
    
    \vspace{2ex}
    
    \centerline{{\scriptsize\bsf B} \raisebox{-0.85\height}{\epsfig{file=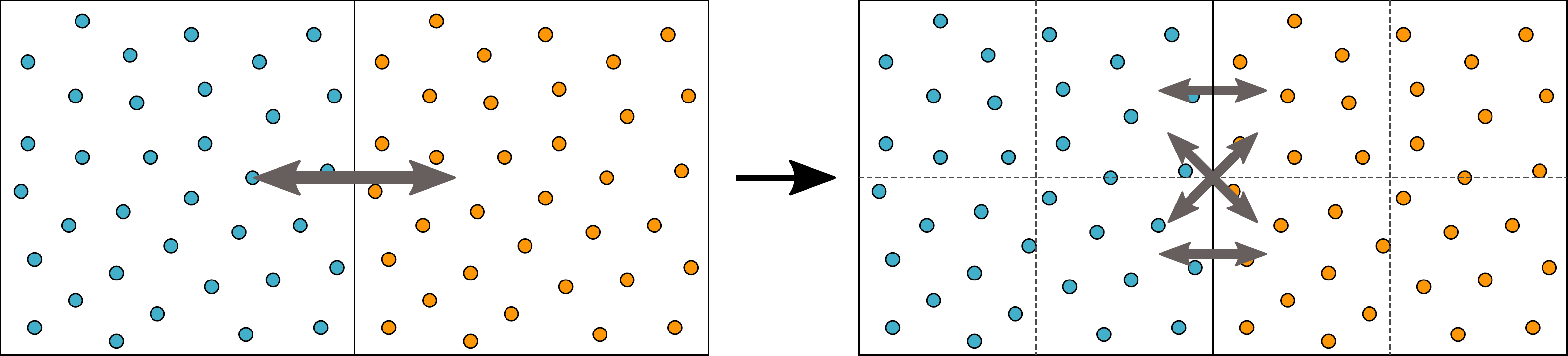,width=0.8\columnwidth}}}
    
    \caption{({\bsf A}) Large cells can be split, and their
        self-interaction replaced by the self- and pair-interactions
        of their sub-cells.
        ({\bsf B}) If all particles in a pair of interacting cells have a smoothing
        length less or equal to half of the cell edge length, both cells can be
        split, and their pair-interaction replaced by the pair-interactions
        of the neighbouring sub-cells across the interface.}
    \label{fig:SplitCell}
    \label{fig:SplitPair}
\end{figure}

After the cells have been split, the cell self-interactions
of each cell can be split up into the self-interaction
of its sub-cells and the pair-interactions between
them.
Likewise, the cell pair-interactions between two cells
that have been split can themselves be split up into
the pair-interactions of the sub-cells spanning the
original pair boundary if, and only if,
all particles in both cells have a smoothing length of
less than half the cell edge length (see \fig{SplitCell}).

If the cells, self-interactions, and pair-interactions are split
in such a way, if two particles are within range of each other,
they will (a) either share a cell for which a cell self-interaction
is defined, or (b) they will be located in two cells which share
a cell pair-interaction.
In order to identify all the particles within range of each other,
it is therefore sufficient to traverse the list of
self-interactions and pair-interactions and inspect the particle
pairs therein.

\subsection{Particle interactions}

The interactions between all particle pairs within a given cell,
i.e. the cell's self-interaction, can be computed by means of
a double {\tt for}-loop over the cell's particle array.
The algorithm, in C-like pseudo code, can be written as follows:

\begin{center}\begin{minipage}{0.9\columnwidth}
    \begin{lstlisting}
for ( i = 0 ; i < count-1 ; i++ ) {
  for ( j = i+1 ; j < count ; j++ ) {
    rij = ||parts[i] - parts[j]||.
    if ( rij < h[i] || rij < h[j] ) {
      compute interaction.
      }
    }
  }
    \end{lstlisting}
\end{minipage}\end{center}

\noindent where {\tt count} is the number of particles in the
cell and {\tt parts} and {\tt h} refers to an array of those
particles and their smoothing lengths respectively.

The interactions between all particles in a pair of cells
can be computed similarly, e.g.:
   
\begin{center}\begin{minipage}{0.9\columnwidth}
    \begin{lstlisting}
for ( i = 0 ; i < count_i ; i++ ) {
  for ( j = 0 ; j < count_j ; j++ ) {
    rij = ||parts_i[i] - parts_j[j]||.
    if ( rij < h_i[i] || rij < h_j[j] ) {
      compute interaction.
      }
    }
  }
    \end{lstlisting}
\end{minipage}\end{center}

\noindent where {\tt count\_i} and {\tt count\_j} refer to
the number of particles in each cell and {\tt parts\_i} and
{\tt parts\_j}, and {\tt h\_i} and {\tt h\_j}, refer to the
particles of each cell and their smoothing lengths respectively.

\begin{figure}
    \centerline{\epsfig{file=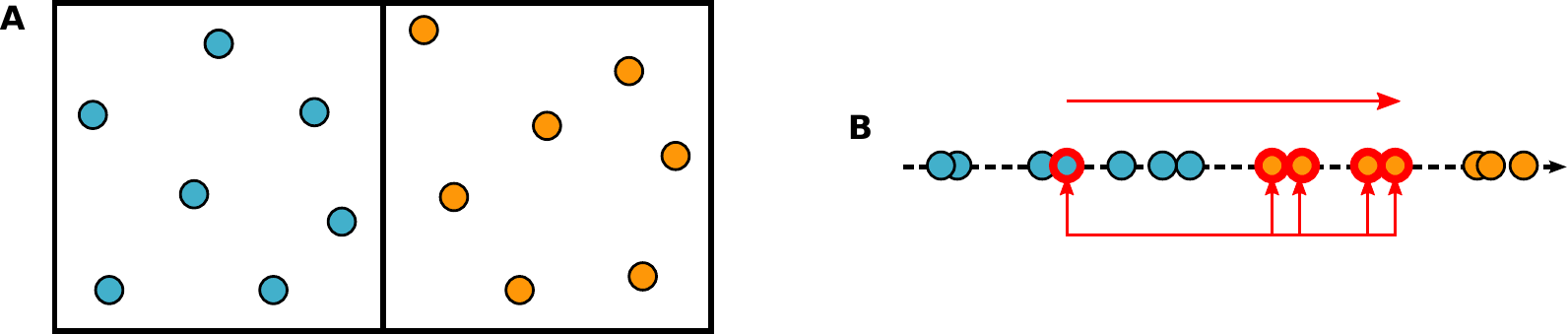,width=0.8\columnwidth}}
    
    \caption{Sorted cell pair-interactions. ({\bsf A}) Starting from a pair of
        neighbouring cells, ({\bsf B}) the particles from both cells
        are projected onto the axis joining the centers of the two cells.
        The particles on the left (blue) and right (orange) are
        then sorted in descending and ascending order respectively.
        Each particle on the left is then only interacted with
        the particles on the right within the cutoff radius along the cell axis.
        }
    \label{fig:SortedInteractions}
\end{figure}

As described in \cite{ref:Gonnet2007}, using this
naive double {\tt for}-loop, for uniform particle
distributions only $33.5\%$, $16.2\%$,
and $3.6\%$ of all particle
pairs between cells sharing a common face, edge, or corner, respectively,
will be within range of each other, leading
to an excessive number of spurious pairwise distance evaluations (line~3).
We will therefore use the sorted cell
interactions described in \cite{ref:Gonnet2007} and \cite{ref:Gonnet2013},
yet with some minor modifications, as
the original algorithm is designed for systems in which the
smoothing lengths of all particles are equal:
We first sort the particles in both cells along the vector joining
the centers of the two cells and then loop over the
parts $p_i$ on the left and interact them with the sorted parts $p_j$
on the right which are within $h_i$ {\em along the cell pair axis}.
The same procedure is repeated for each particle $p_j$ on the
right, interacting with each other particle $p_i$ on the
left, which is within $h_j$, {\em but not within} $h_i$, along
the cell pair axis (see \fig{SortedInteractions}).
The resulting algorithm, in C-like pseudo-code, can be written as follows:
        
\begin{center}\begin{minipage}{0.9\columnwidth}
    \begin{lstlisting}
for ( i = 0 ; i < count_i ; i++ ) {
  for ( jj = 0 ; jj < count_j ; jj++ ) {
    j = ind_j[jj];
    if ( r_i[i] + h_i[i] < r_j[j] )
      break;
    rij = ||parts_i[i] - parts_j[j]||.
    if ( rij < h_i[i] ) {
      compute interaction.
      }
    }
  }
for ( j = 0 ; j < count_j ; j++ ) {
  for ( ii = count_i-1 ; ii >= 0 ; ii-- ) {
    i = ind_i[i];
    if ( r_i[i] < r_j[j] - h_j[j] )
      break;
    rij = ||parts_i[i] - parts_j[j]||.
    if ( rij < h_j[j] && rij > h_i[i] ) {
      compute interaction.
      }
    }
  }
    \end{lstlisting}
\end{minipage}\end{center}
        
\noindent where {\tt r\_i} and {\tt r\_j} contains the position
the particles of both cells along the cell axis, and
{\tt ind\_i} and {\tt ind\_j} contain the particle indices
sorted with respect to these positions respectively.

This may seem like quite a bit of sorting, but
if a cell has been split and its sub-cells have been sorted,
the sorted indices of the higher-level cell can be constructed
by shifting and merging the indices of its eight sub-cells.
This reduces the \oh{n\log{n}} sorting to \oh{n} for merging. 
Furthermore, instead of sorting the particles every time we compute the
pairwise interactions between two cells, we can pre-compute
the sorted indices along the 13 possible cell-pair axes and
store them for each cell, i.e. as is done in \cite{ref:Gonnet2013}.
The sorted indices can also be re-used over several time steps
until the hierarchical cell decomposition needs to be
re-computed.

\subsection{Parallel implementation}

\begin{figure}
    \centerline{\epsfig{file=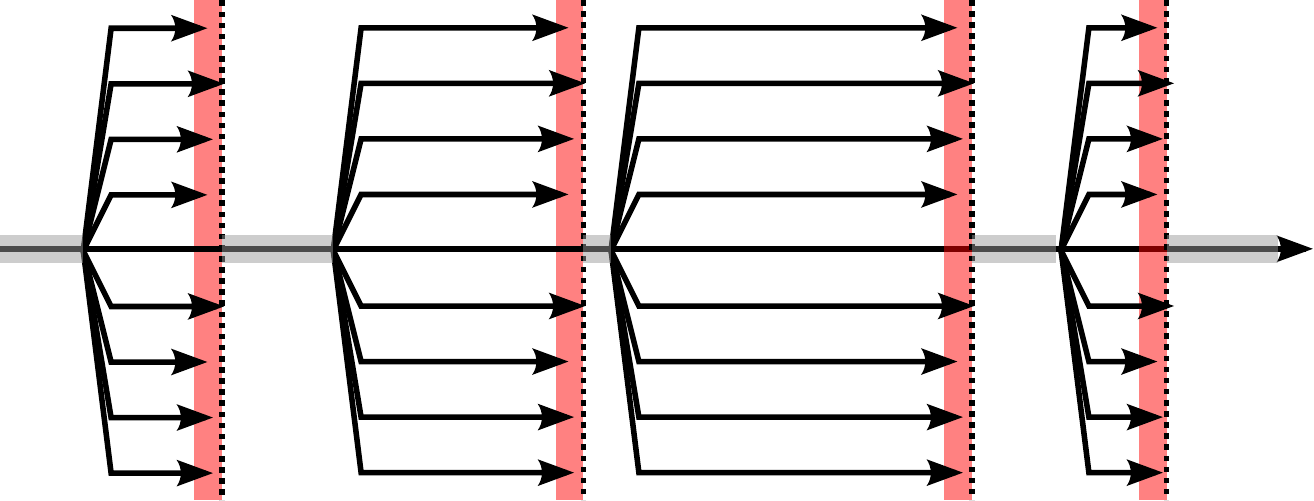,width=0.8\columnwidth}}
    
    \caption{Branch-and-bound parallelism as is commonly used in OpenMP.
        The horizontal arrows indicate the program flow over time, and
        branching arrows indicate a parallel section. The dotted vertical
        bars are the synchronization points at the end of each such section.
        Parallel efficiency is lost due to two factors: The grey shaded areas
        along the main horizontal area indicate parts of the program that
        do not execute in parallel and restrict the maximum parallel
        efficiency, e.g.~as described by Amdahl's law, and the red
        areas indicate the difference between the fastest and slowest
        threads in each parallel block, i.e.~the time lost to
        thread synchronization.
        }
    \label{fig:OMPScaling}
\end{figure}

OpenMP \cite{ref:Dagum1998} is arguably the most well-known
paradigm for shared-memory, or multi-threaded parallelism.
In OpenMP, compiler annotations are used to describe if and when
specific loops or portions of the code can be executed
in parallel.
When such a parallel section, e.g.~a parallel loop, is
encountered, the sections of the loop are split statically
or dynamically over the available threads, each executing
on a single core, and merge at the end of the parallel section.
Unfortunately, this can lead to a lot of inefficient
branch-and-bound type operations, which generally lead to
low performance and bad scaling on even moderate numbers
of cores (see \fig{OMPScaling}).

Furthermore, this form of shared-memory parallelism provides
no implicit mechanism to avoid or handle concurrency problems,
e.g.~two threads attempting to modify the same data at the same time,
or data dependencies between them.
These must be implemented explicitly using either redundancy, barriers,
critical sections, or atomic memory operations, which can further degrade
parallel performance.

In order to better exploit shared-memory parallelism, 
we have to change the underlying paradigm, i.e. instead
of annotating an essentially serial computation with parallel
bits, it is preferable to describe the whole computation in a way that
is inherently parallelizable.
One such approach is {\em task-based parallelism}, in which the
computation is divided into a number of inter-dependent
computational tasks, which are then dynamically allocated
to a number of processors.
In order to ensure that the tasks are executed in the right
order, e.g.~that data needed by one task is only used once it
has been produced by another task, and that no two tasks
update the same data at the same time, constraints are specified
and enforced by the task scheduler.

Several middle-wares providing such task-based
parallelism exist, e.g.~Cilk \cite{ref:Blumofe1995}, QUARK \cite{ref:QUARK},
StarPU \cite{ref:Augonnet2011}, SMP~Superscalar \cite{ref:SMPSuperscalar},
OpenMP~3.0 \cite{ref:Duran2009}, and Intel's TBB \cite{ref:Reinders2007}.
These implementations allow for the specification of individual tasks
along with their {\em dependencies}, i.e.~hierarchical relationships
specifying which tasks must have completed before another task
can be executed.

We will, however, differ from these approaches in that we introduce the
concept of {\em conflicts} between tasks.
Conflicts occur when two tasks operate on the same data, 
but the order in which these operations must occur is not defined.
In previous task-based models, conflicts can be modeled as dependencies,
yet this introduces an artificial ordering between the tasks
and imposes unnecessary constraints on the task scheduler
(e.g.~mutual and non-mutual interactions in \cite{ref:Ltaief2012}).
These conflicts can be modelled using exclusive {\em locks} on shared
resources, i.e.~a task will only be scheduled if it can obtain an
exclusive lock on, and thus exclusive access to, potentially shared data.

The particle interactions described in the previous subsection
lead to three basic task types:
\begin{itemize}
    \item Cell {\em sorting}, in which the particles in a given
        cell are sorted with respect to their position along a
        one-dimensional axis,
    \item Cell {\em self-interaction}, in which all the particles
        of a given cell are interacted with all the other particles
        within the same cell,
    \item Cell {\em pair-interaction}, in which the interactions for
        all particle pairs spanning a pair of cells are computed. 
\end{itemize}
The self-interaction and pair-interaction tasks exist in
two flavors, one for the density computation
and one for the force computation.
Each pair-interaction task requires the sorted indices of
the particles in each cell provided by the sorting tasks.
Since the tasks are restricted to operating on the data of a
single cell, or pair of cells, two tasks conflict if they
operate on overlapping sets of cells.
Due to the hierarchical nature of the spatial decomposition,
two tasks also conflict if the cells of one task are sub-cells
of the other.

Since the SPH computations have two phases, i.e.~density and force
computation, we introduce a {\em ghost} task in between for each cell.
This ghost task depends on all the density computations
for a given cell, and, in turn, all force computations involving
that cell depend on its ghost task.
Using this mechanism, we can enforce that all density computations
for a set of particles have completed before we use this
density in the force computations.

The dependencies and conflicts between tasks can thus be
formulated as follows:

\begin{itemize}

    \item A cell sorting task on a cell with sub-cells depends
        on the sorting tasks of all its sub-cells.

    \item A cell pair-interaction depends on the cell sorting
        tasks of both its cells.
        
    \item Cell pair-interaction and cell self-interaction tasks
        operating on overlapping sets of cells or sub-cells
        conflict with each other.
        
    \item The ghost task of each cell depends on all the density cell pair
        interactions and self-interactions which involve the particles
        in that cell.
        
    \item The ghost task of each cell depends on the ghost tasks of
        its sub-cells.
        
    \item The force cell pair-interaction and self-interaction tasks
        depend on the ghost tasks of the cells on which they operate.

\end{itemize}

\begin{figure}
    \centerline{\epsfig{file=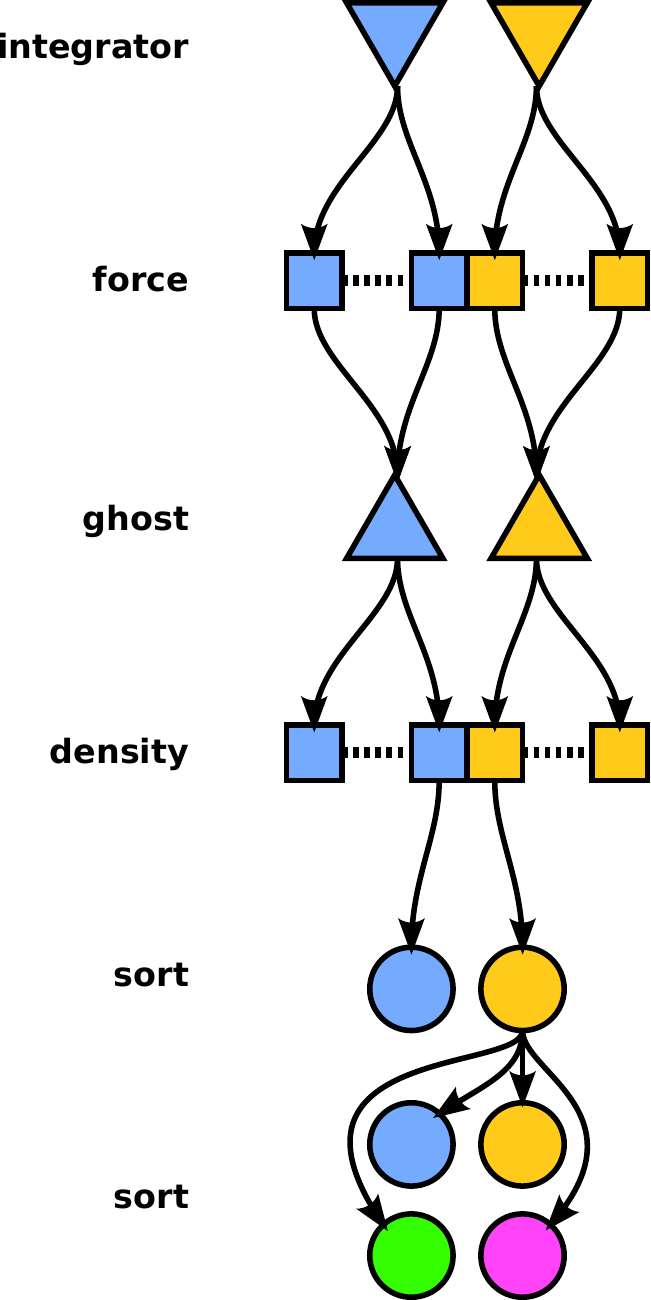,width=0.4\columnwidth}}
    
    \caption{Task dependencies and conflicts:
        Arrows indicate the dependencies
        between different task types, i.e.~and arrow from task A to task
        B indicates that A depends on B.
        Horizontal dashed lines between tasks indicate conflicts,
        i.e.~the two tasks can not be executed concurrently.
        Each sort task (circles) depends
        only on the sort tasks of its sub-cells.
        The pair-interactions (rectangles) for the particle
        density computation depend on the sort tasks of the respective cells,
        whereas self-interaction tasks (squares) for the density computation
        do not, as they do not require sorting.
        Self- and pair-Interactions on overlapping cells (same colour)
        conflict with each other.
        The ghost task of each cell (triangles) depends on the self-
        and pair-interaction density tasks.
        The self- and pair-interaction tasks for the force computation,
        finally, depend on the ghost tasks of the respective cells.
        }
    \label{fig:Hierarchy2}
\end{figure}

\noindent These task dependencies are illustrated in \fig{Hierarchy2}.

This task decomposition has significant advantages over the use of
spatial trees:
First of all, the cost of identifying all particles in range of
a given particle does not depend on the total number of
particles.
Furthermore, the particle interactions in each task are computed
symmetrically, i.e.~each particle pair is identified only
once for each interaction type.
The sorted particle indices can be re-used for both the
density and force computation, and over several time-steps,
thus reducing the computational cost even further.
Finally, if the particles are stored grouped by cell, each task
then only involves accessing and modifying a contained
and contiguous regions of memory, thus potentially improving
cache re-use \cite{ref:Fomin2011}.

%%%%%%%%%%%%%%%%%%%%%%%%%%%%%%%%%%%%%%%%%%%%%%%%%%%%%%%%%%%%%%%%%%%%%%%%%%%%%%%%
%  Validation of the algorithms
%%%%%%%%%%%%%%%%%%%%%%%%%%%%%%%%%%%%%%%%%%%%%%%%%%%%%%%%%%%%%%%%%%%%%%%%%%%%%%%%
\section{Validation}

In the following, we describe how the algorithms shown in the previous section
are implemented and tested against exiting codes on specific test problems.

\subsection{Implementation details}

The algorithms described above are all implemented as part
of \swift (\underline{S}PH \underline{W}ith
\underline{I}nter-dependent \underline{F}ine-grained
\underline{T}asking),
an Open-Source platform for hybrid shared/distributed-memory
SPH simulations\footnote{See http://swiftsim.sourceforge.net/}.

\swift is implemented in C, and can be compiled with the
{\tt gcc} compiler.
Although SIMD-vectorized code, using the {\tt gcc} vector types
and SSE/AVX intrinsics, has been implemented, it was switched
off in the following to allow for a fair comparison against
unvectorized codes.

The code for the task-based parallelism is implemented using
standard {\tt pthread}s \cite{ref:pthreads}, and in some places,
e.g.~in the time-stepper
or task list creation, OpenMP is used.
Each thread is assigned its own task queue, over which the tasks
are distributed evenly in topological order of the dependencies.
If a thread runs out of tasks, it can steal from another task's
queue \cite{ref:Blumofe1999}.
The threads are initialized once at the start of the simulation
and synchronize via a barrier between time steps.

\subsection{Simulation setup}

In order to test their accuracy, efficiency, and scaling,
the algorithms described in the previous section
were tested in \swift using the following three
simulation setups:
\begin{itemize}
    \item {\em Sod-shock} \cite{ref:Sod1978}: A cubic periodic domain containing a
        high-density region of
        800\,000 particles with $P_i=1$ and $\rho_i=1$ on one half,
        and a low-density region of 200\,000
        particles with $P_i=0.1795$ and $\rho_i=1/4$ on the other.
    \item {\em Sedov blast}: A cubic grid of $101\times 101\times 101$
        particles at rest with $P_i=10^{-5}$ and $\rho_i=1$, yet with the
        energy of the central 33 particles set to
        $u_i=100/33/m_i$.
    \item {\em Cosmological volume}: Realistic distribution of matter
        in a periodic volume of universe. The simulation consists
        of 1\,841\,127 particles with
        a mix of smoothing lengths spanning three orders of magnitude,
        providing a test-case for neighbour finding and parallel
        scaling in a real-world scenario.
\end{itemize}
The first two cases are used both for benchmarking and for
validating the correctness of the code.
In all simulations, the constants $N_{ngb}=48$, $\gamma=5/3$,
$C_{CFL}=0.3$, and $\alpha = 2.0$ (viscosity) were used.
In \swift, cells were split if they contained more than 400
particles and more than 87.5\% of the particles had a smoothing
length less than half the cell edge length.

In all three test cases, results were
computed using a fixed time step and updating the forces on all
particles in each time-step.

The simulation results were compared with Gadget-2 \cite{ref:Springel2005}
in terms of speed and parallel scaling.
Gadget-2 was compiled with the Intel C Compiler version 2013.0.028
using the options \texttt{\small -DUSE\_IRECV -O3 -ip -fp-model fast -ftz
-no-prec-div -mcmodel=medium}.
For the parallel runs, OpenMPI version 1.6.3 was used \cite{ref:Gabriel2004}.

\swift v.~0.1.0 was compiled with the GNU C Compiler version 4.7.2
using the options \texttt{\small -O3 -ffast-math -fstrict-aliasing
-ftree-vectorize -funroll-loops -mmmx -msse -msse2 -msse3 -mssse3
-msse4.1 -msse4.2 -mavx -fopenmp -march=native -pthread}.
Note that although the compiler switches for the SSE and AVX
vector instruction sets were activated, explicit
SIMD-vectorization was not used.

All simulations were run on a $4\times$Intel Xeon E5-4640
32-core machine running at 2.4\,GHz with
CentOS release 6.2 Linux for x86\_64.

\subsection{Results}

\begin{figure*}
    \centerline{\epsfig{file=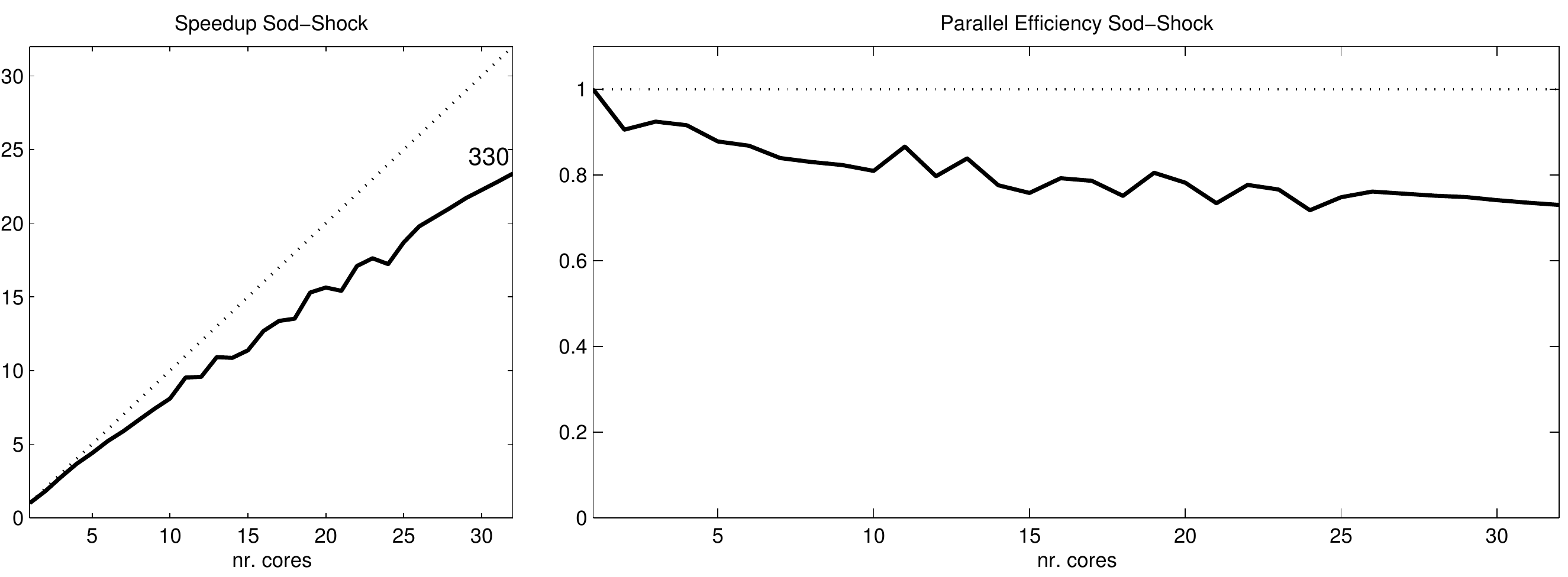,width=0.8\textwidth}}
    \centerline{\epsfig{file=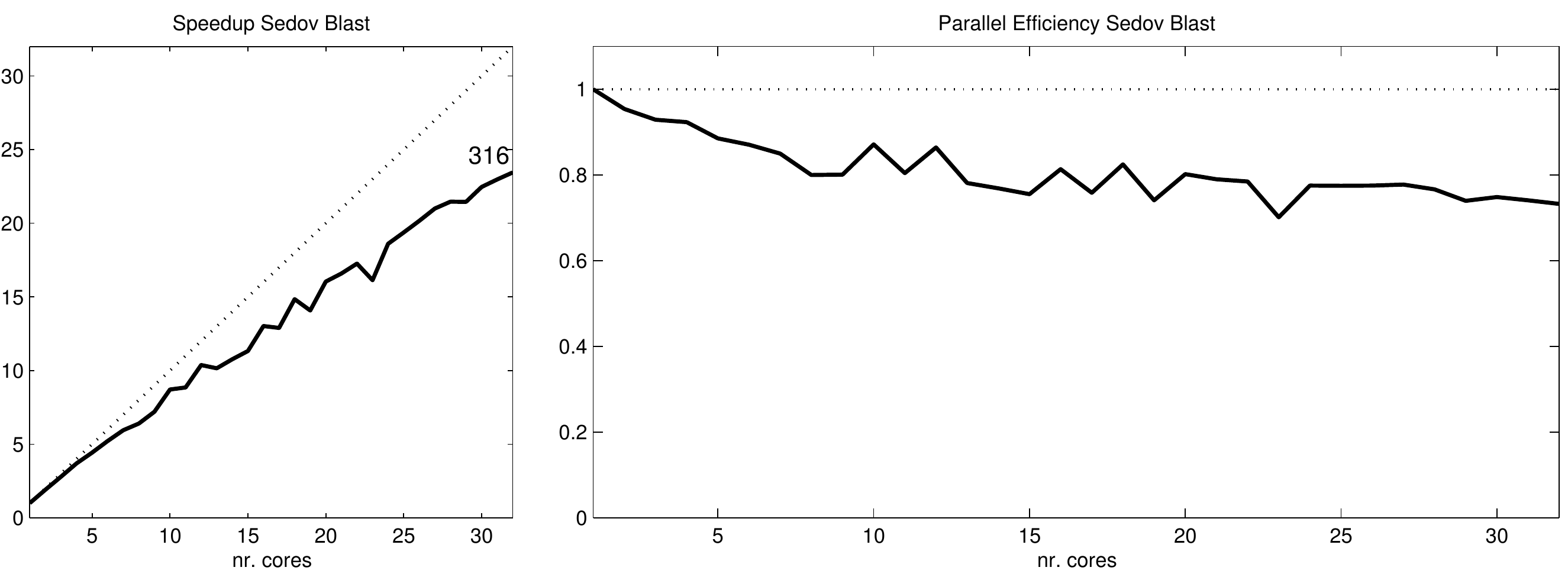,width=0.8\textwidth}}
    \centerline{\epsfig{file=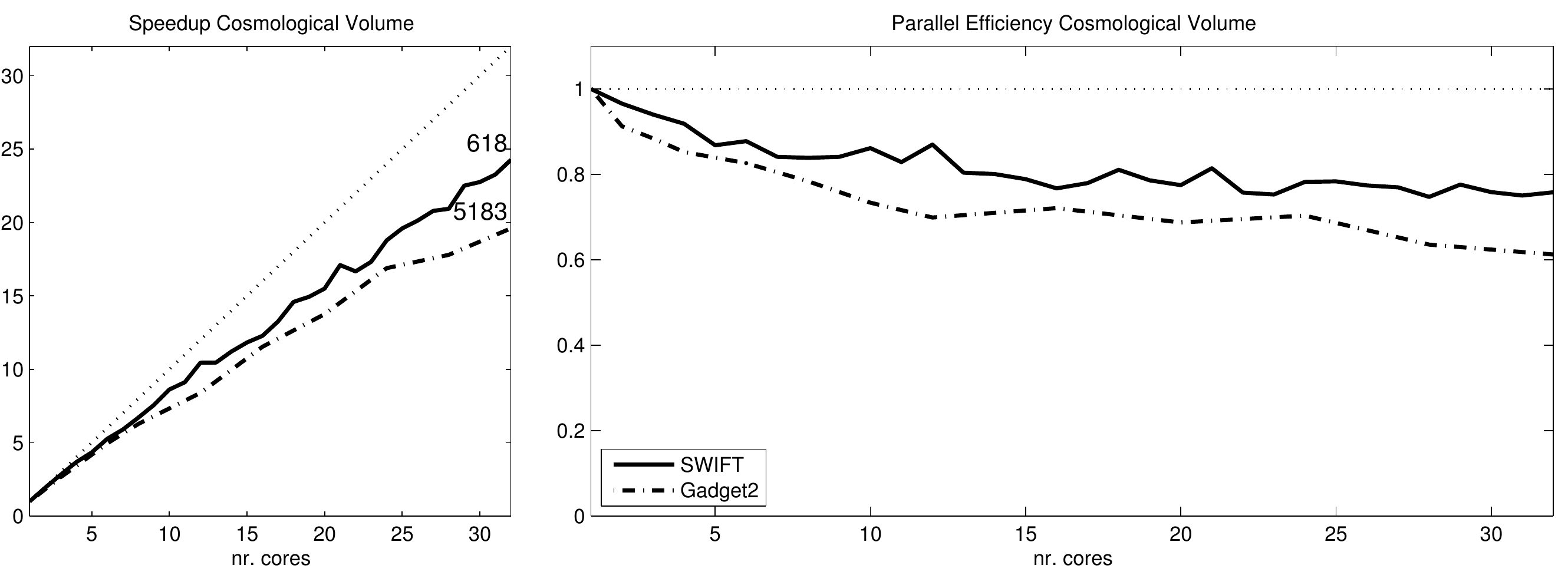,width=0.8\textwidth}}
    
    \caption{Parallel scaling and efficiency for the Sod-shock, Sedov blast
        Cosmological volume simulations.
        The numbers in the scaling plots indicate the average milliseconds
        per time step when running on all 32 cores.}
    \label{fig:Results}
\end{figure*}

The scaling and efficiency of \swift for the simulations described
in the previous subsection are shown in \fig{Results}.
In all three cases, \swift obtains a parallel efficiency of over
75\% on 32 cores.
For the Cosmological volume, it is also more than eight times
faster than Gadget-2.

\begin{figure}
    \centerline{\epsfig{file=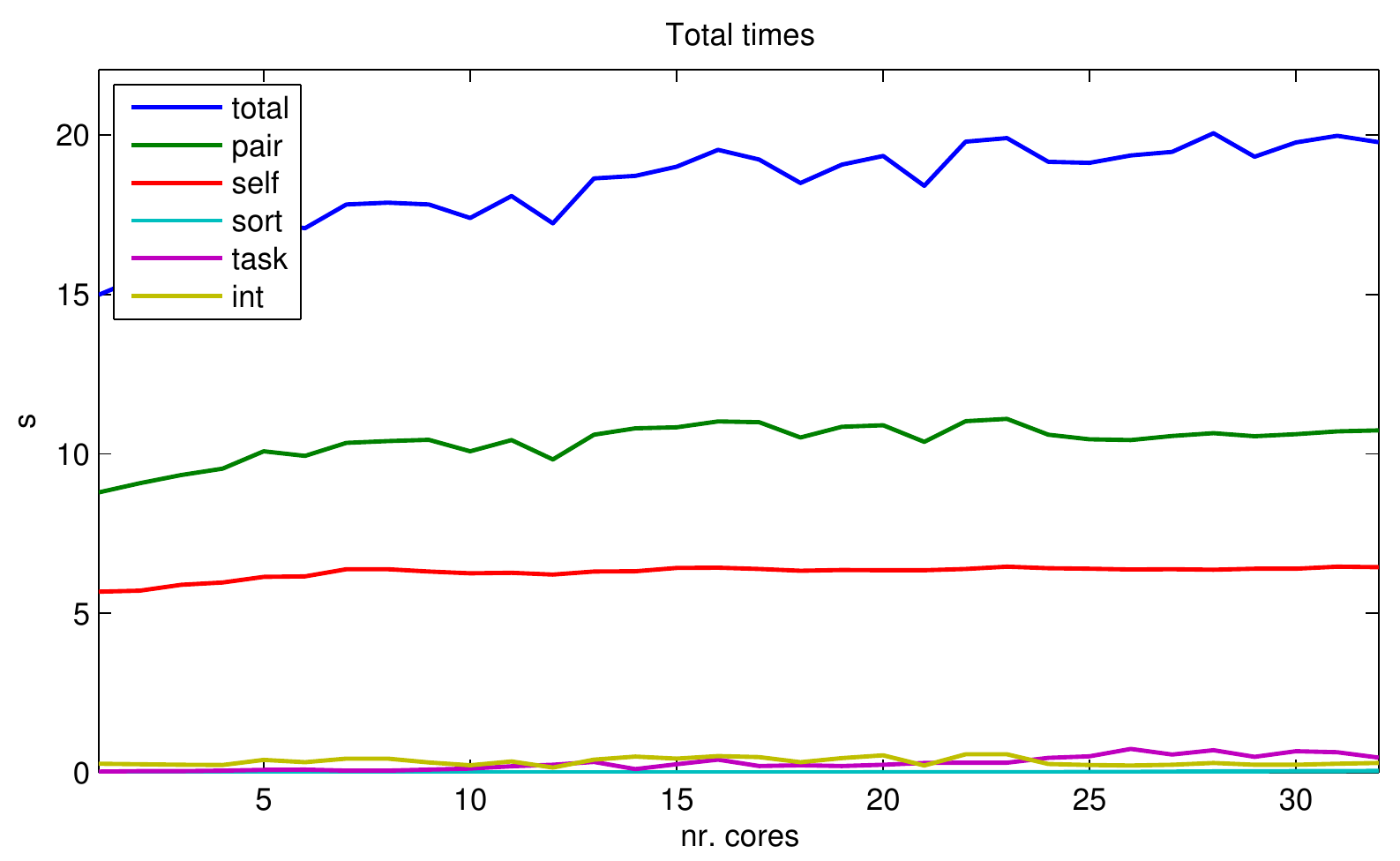,width=0.9\columnwidth}}
    
    \caption{Total time for each task type, plus overheads for task
        management and time integration.
        The values, in seconds, are summed over all processors.}
    \label{fig:Times}
\end{figure}

\fig{Times} shows the total contribution of each task type, as well
as the overheads for task acquisition and time integration
for the Cosmological volume simulation.
Note that efficiency is lost principally in the self and pair 
interactions, which is most probably due to the reduction of
the effective memory bandwidth with the increasing number of cores
used.
The overheads for task acquisition contribute less than 3.5\% to
the total costs.

\begin{figure}
    \centerline{\epsfig{file=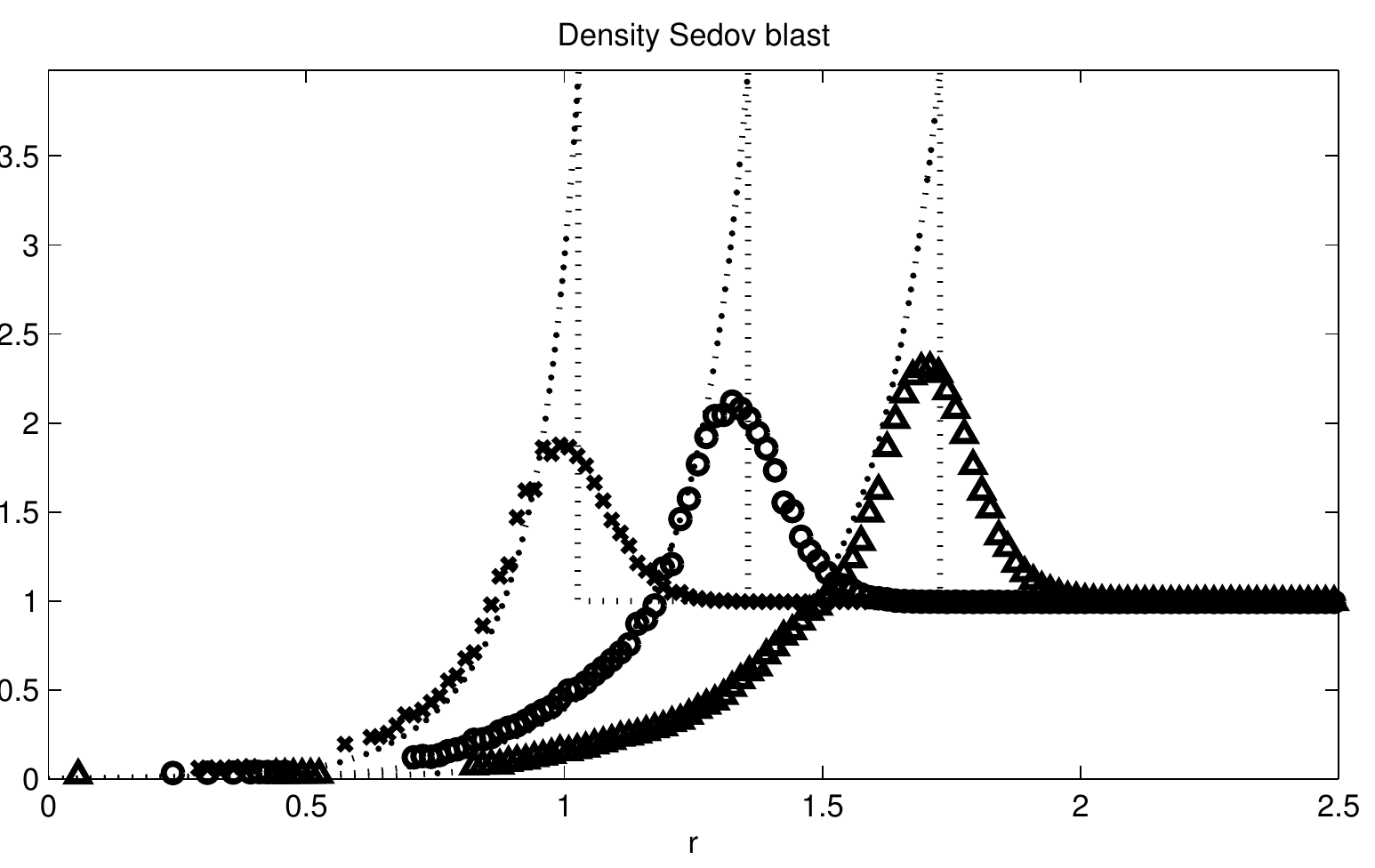,width=0.9\columnwidth}}
    
    \caption{Average radial density profiles and analytical solutions
        (dotted lines)
        for the Sedov blast simulation at
        times $t=\{0.075, 0.15, 0.275\}$.}
    \label{fig:Sedov_density}
\end{figure}

\begin{figure*}
    \centerline{\epsfig{file=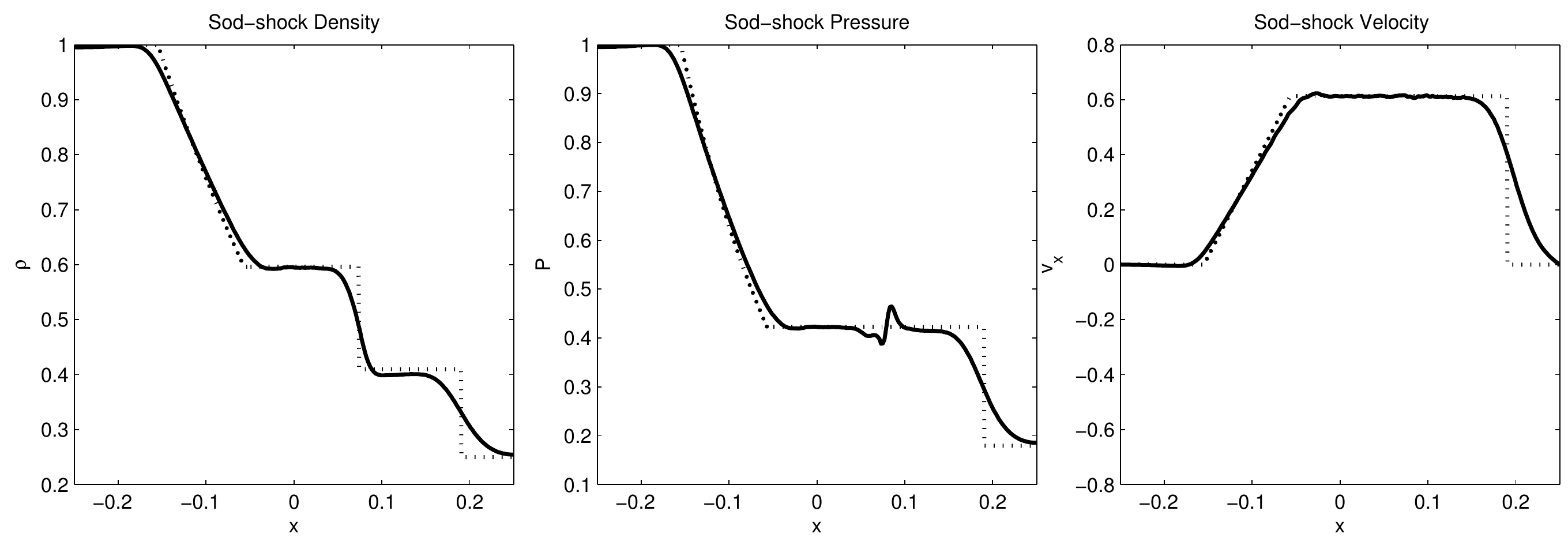,width=0.8\textwidth}}
    
    \caption{Average density, pressure, and velocity in $x$-direction,
        along with the corresponding analytical solutions (dotted lines)
        for the Sod-shock simulations at $t=0.12$.}
    \label{fig:SodShock_profiles}
\end{figure*}

In order to assess the correctness of the simulation, average radial
density, pressure, and velocity profiles were computed for
the Sedov blast (see \fig{Sedov_density}) and Sod-shock
(see \fig{SodShock_profiles}) simulations and compared with
the analytical results given in \cite{ref:Sedov1959} and
\cite{ref:Sod1978} respectively.
These resutls are comparable to the output produced by
Gadget-2 for the same initial conditions, and sharper
results could be achieved at higher resolutions and/or
different parameterizations.

%%%%%%%%%%%%%%%%%%%%%%%%%%%%%%%%%%%%%%%%%%%%%%%%%%%%%%%%%%%%%%%%%%%%%%%%%%%%%%%%
%  Conclusions
%%%%%%%%%%%%%%%%%%%%%%%%%%%%%%%%%%%%%%%%%%%%%%%%%%%%%%%%%%%%%%%%%%%%%%%%%%%%%%%%
\section{Conclusions}

The results show that \swift is significantly more efficient than
current state-of-the-art tree-based methods.
The improved performance is the result of several features:
\begin{itemize}
    \item Cell-based neighbour searching which requires only
        constant time per particle,
    \item Sorted particle interactions which significantly
        reduce the number of spurious pairwise distance
        computations in the neighbour search,
    \item Use of symmetry in computing the particle densities
        and forces, thus requiring each particle pair to be
        found only once,
    \item Better cache efficiency of the cell-based interactions.
\end{itemize}
Since no hardware-specific tricks or explicit
SIMD-vectorization were used, although the latter is available
in the code, they can not be made responsible for the
improved efficiency.

In addition to the increased speed, \swift
also shows good scaling, in excess
of 75\% parallel efficiency on 32 cores, for all three simulations.
This is due to the fine-grained load balancing and low number
of synchronization points as a result of the task-based parallelism scheme.

We are currently working to extend \swift for hybrid
shared/distributed-memory parallelism, and to GPUs
\cite{ref:Chalk2013}.
Task-based parallelism will play an important role for
both these developments: In hybrid schemes, asynchronous communication
between distributed-memory nodes can be modelled as simply another
set of tasks with their corresponding dependencies, thus avoiding
excessive synchronization points or latencies.
As has already been shown for molecular dynamics simulations,
the task-based parallel scheme can also easily be extended
for computations on GPUs, using the same set of tasks and 
dependencies.
Work is also ongoing in adding different and better implementations
of particle-based hydrodynamics, i.e.~for cosmological
simulations \cite{ref:Schaller2013}.

%%%%%%%%%%%%%%%%%%%%%%%%%%%%%%%%%%%%%%%%%%%%%%%%%%%%%%%%%%%%%%%%%%%%%%%%%%%%%%%%
%  Acknowledgments
%%%%%%%%%%%%%%%%%%%%%%%%%%%%%%%%%%%%%%%%%%%%%%%%%%%%%%%%%%%%%%%%%%%%%%%%%%%%%%%%
\section*{Acknowledgments}

The authors would like to thank Lydia Heck of the Institute for Computational
Cosmology, Durham University, for having initiated this interdisciplinary
collaboration, and for providing exclusive access to the hardware on which
all benchmarks were run.



% http://www.ctan.org/tex-archive/macros/latex/contrib/supported/IEEEtran/bibtex
\bibliographystyle{IEEEtran.bst}
\bibliography{sph.bib}
%

\end{document}